\documentstyle[twoside]{article}
\catcode`\@=11
\long\def\@makefntext#1{
\protect\noindent \hbox to 3.2pt {\hskip-.9pt
$^{{\eightrm\@thefnmark}}$\hfil}#1\hfill}               

\def\thefootnote{\fnsymbol{footnote}}
\def\@makefnmark{\hbox to 0pt{$^{\@thefnmark}$\hss}}    

\def\ps@myheadings{\let\@mkboth\@gobbletwo
\def\@oddhead{\hbox{}
\rightmark\hfil\eightrm\thepage}
\def\@oddfoot{}\def\@evenhead{\eightrm\thepage\hfil
\leftmark\hbox{}}\def\@evenfoot{}
\def\sectionmark##1{}\def\subsectionmark##1{}}

\oddsidemargin=\evensidemargin
\renewcommand{\thefootnote}{\fnsymbol{footnote}}

\newcounter{sectionc}\newcounter{subsectionc}\newcounter{subsubsectionc}
\renewcommand{\section}[1] {\vspace{12pt}\addtocounter{sectionc}{1}
\setcounter{subsectionc}{0}\setcounter{subsubsectionc}{0}\noindent
        {\tenbf\thesectionc. #1}\par\vspace{5pt}}
\renewcommand{\subsection}[1] {\vspace{12pt}\addtocounter{subsectionc}{1}
        \setcounter{subsubsectionc}{0}\noindent
        {\bf\thesectionc.\thesubsectionc. {\kern1pt \bfit #1}}\par\vspace{5pt}}
\renewcommand{\subsubsection}[1] {\vspace{12pt}\addtocounter{subsubsectionc}{1}
        \noindent{\tenrm\thesectionc.\thesubsectionc.\thesubsubsectionc.
        {\kern1pt \tenit #1}}\par\vspace{5pt}}

\newcounter{appendixc}
\newcounter{subappendixc}[appendixc]
\newcounter{subsubappendixc}[subappendixc]
\renewcommand{\thesubappendixc}{\Alph{appendixc}.\arabic{subappendixc}}
\renewcommand{\thesubsubappendixc}
        {\Alph{appendixc}.\arabic{subappendixc}.\arabic{subsubappendixc}}

\renewcommand{\appendix}[1] {\vspace{12pt}
        \refstepcounter{appendixc}
        \setcounter{figure}{0}
        \setcounter{table}{0}
        \setcounter{lemma}{0}
        \setcounter{theorem}{0}
        \setcounter{corollary}{0}
        \setcounter{definition}{0}
        \setcounter{equation}{0}
        \renewcommand{\thefigure}{\Alph{appendixc}.\arabic{figure}}
        \renewcommand{\thetable}{\Alph{appendixc}.\arabic{table}}
        \renewcommand{\theappendixc}{\Alph{appendixc}}
        \renewcommand{\thelemma}{\Alph{appendixc}.\arabic{lemma}}
        \renewcommand{\thetheorem}{\Alph{appendixc}.\arabic{theorem}}
        \renewcommand{\thedefinition}{\Alph{appendixc}.\arabic{definition}}
        \renewcommand{\thecorollary}{\Alph{appendixc}.\arabic{corollary}}
        \renewcommand{\theequation}{\Alph{appendixc}.\arabic{equation}}
        \noindent{\tenbf Appendix \theappendixc #1}\par\vspace{5pt}}
\newcommand{\subappendix}[1] {\vspace{12pt}
        \refstepcounter{subappendixc}
        \noindent{\bf Appendix \thesubappendixc. {\kern1pt \bfit #1}}
        \par\vspace{5pt}}
\newcommand{\subsubappendix}[1] {\vspace{12pt}
        \refstepcounter{subsubappendixc}
        \noindent{\rm Appendix \thesubsubappendixc. {\kern1pt \tenit #1}}
        \par\vspace{5pt}}

\newcommand{\textlineskip}{\baselineskip=13pt}
\newcommand{\smalllineskip}{\baselineskip=10pt}

\def\eightcirc{
\begin{picture}(0,0)
\put(4.4,1.8){\circle{6.5}}
\end{picture}}
\def\eightcopyright{\eightcirc\kern2.7pt\hbox{\eightrm c}}

\newcommand{\copyrightheading}[1]
        {\vspace*{-2.5cm}\smalllineskip{\flushleft
        {\footnotesize International Journal of Modern Physics  A }\\
        {\footnotesize $\eightcopyright$\, World Scientific Publishing
         Company}\\
         }}

\renewenvironment{thebibliography}[1]
        {\frenchspacing
         \ninerm\baselineskip=11pt
         \begin{list}{\arabic{enumi}.}
        {\usecounter{enumi}\setlength{\parsep}{0pt}
         \setlength{\leftmargin 12.7pt}{\rightmargin 0pt} 
         \setlength{\itemsep}{0pt} \settowidth
        {\labelwidth}{#1.}\sloppy}}{\end{list}}

\newcounter{itemlistc}
\newcounter{romanlistc}
\newcounter{alphlistc}
\newcounter{arabiclistc}

\newcommand{\fcaption}[1]{
        \refstepcounter{figure}
        \setbox\@tempboxa = \hbox{\footnotesize Fig.~\thefigure. #1}
        \ifdim \wd\@tempboxa > 5in
           {\begin{center}
        \parbox{5in}{\footnotesize\smalllineskip Fig.~\thefigure. #1}
            \end{center}}
        \else
             {\begin{center}
             {\footnotesize Fig.~\thefigure. #1}
              \end{center}}
        \fi}

\newcommand{\tcaption}[1]{
        \refstepcounter{table}
        \setbox\@tempboxa = \hbox{\footnotesize Table~\thetable. #1}
        \ifdim \wd\@tempboxa > 5in
           {\begin{center}
        \parbox{5in}{\footnotesize\smalllineskip Table~\thetable. #1}
            \end{center}}
        \else
             {\begin{center}
             {\footnotesize Table~\thetable. #1}
              \end{center}}
        \fi}

\def\@citex[#1]#2{\if@filesw\immediate\write\@auxout
        {\string\citation{#2}}\fi
\def\@citea{}\@cite{\@for\@citeb:=#2\do
        {\@citea\def\@citea{,}\@ifundefined
        {b@\@citeb}{{\bf ?}\@warning
        {Citation `\@citeb' on page \thepage \space undefined}}
        {\csname b@\@citeb\endcsname}}}{#1}}

\newif\if@cghi
\def\cite{\@cghitrue\@ifnextchar [{\@tempswatrue
        \@citex}{\@tempswafalse\@citex[]}}
\def\citelow{\@cghifalse\@ifnextchar [{\@tempswatrue
        \@citex}{\@tempswafalse\@citex[]}}
\def\@cite#1#2{{$\null^{#1}$\if@tempswa\typeout
        {IJCGA warning: optional citation argument
        ignored: `#2'} \fi}}

\def\@refcitex[#1]#2{\if@filesw\immediate\write\@auxout
        {\string\citation{#2}}\fi
\def\@citea{}\@refcite{\@for\@citeb:=#2\do
        {\@citea\def\@citea{, }\@ifundefined
        {b@\@citeb}{{\bf ?}\@warning
        {Citation `\@citeb' on page \thepage \space undefined}}
        \hbox{\csname b@\@citeb\endcsname}}}{#1}}

\def\@refcite#1#2{{#1\if@tempswa\typeout
        {IJCGA warning: optional citation argument
        ignored: `#2'} \fi}}

\def\refcite{\@ifnextchar[{\@tempswatrue
        \@refcitex}{\@tempswafalse\@refcitex[]}}

\def\pmb#1{\setbox0=\hbox{#1}
        \kern-.025em\copy0\kern-\wd0
        \kern.05em\copy0\kern-\wd0
        \kern-.025em\raise.0433em\box0}

\def\fnm#1{$^{\mbox{\scriptsize #1}}$}
\def\fnt#1#2{\footnotetext{\kern-.3em
        {$^{\mbox{\scriptsize #1}}$}{#2}}}

\def\fpage#1{\begingroup
\voffset=.3in
\thispagestyle{empty}\begin{table}[b]\centerline{\footnotesize #1}
        \end{table}\endgroup}

\def\runninghead#1#2{\pagestyle{myheadings}
\markboth{{\protect\footnotesize\it{\quad #1}}\hfill}
{\hfill{\protect\footnotesize\it{#2\quad}}}}
\font\tenrm=cmr10
\font\tenit=cmti10
\font\tenbf=cmbx10
\font\bfit=cmbxti10 at 10pt
\font\ninerm=cmr9

\font\eightrm=cmr8

\def\qed{\hbox{${\vcenter{\vbox{                        
   \hrule height 0.4pt\hbox{\vrule width 0.4pt height 6pt
   \kern5pt\vrule width 0.4pt}\hrule height 0.4pt}}}$}}

\renewcommand{\thefootnote}{\fnsymbol{footnote}}        

\begin{document}

\def\be{\begin{equation}}
\def\ee{\end{equation}}


\runninghead{M. Kirchbach}
{Classifying Reported and ``Missing'' Resonances$\ldots$}
\normalsize\textlineskip
\thispagestyle{empty}
\setcounter{page}{1}

\copyrightheading{}                     

\vspace*{0.88truein}

\fpage{1}


\centerline{CLASSIFYING REPORTED AND `MISSING' RESONANCES ACCORDING}
\vspace*{0.075truein}
\centerline{TO THEIR $\bf P$ AND $\bf C$ PROPERTIES}

\vspace*{0.37truein}
\centerline{\footnotesize M. KIRCHBACH\fnm{*}\fnt{*}
{E-mail: mariana@kph.uni-mainz.de; kirchbach@chiral.reduaz.mx }
}
\vspace*{0.015truein}
\baselineskip=10pt
\centerline{\footnotesize\it Escuela de Fisica}
\centerline{\footnotesize\it  
Univ.\ Aut. de Zacatecas}
\centerline{\footnotesize\it Apartado Postal C-580}
\centerline{\footnotesize\it Zacatecas. ZAC 98068 Mexico}
\centerline{\footnotesize\it and}
\centerline{\footnotesize\it Institut f\"ur Kernphysik}
\centerline{\footnotesize\it  
Universit\"at Mainz}
\centerline{\footnotesize\it D-55099 Mainz, Germany}

\vspace*{0.015truein}
\bigskip
\begin{abstract}
The Hilbert space ${\cal H}^{3q}$ of the three quarks with one excited
quark is decomposed into Lorentz group representations. It is shown
that the quantum numbers of the reported and ``missing'' resonances
fall apart and populate distinct representations that differ by their
parity or/and charge conjugation properties. In this way, reported and
\hbox{``missing''} resonances become distinguishable. For example,
resonances from the full listing \hbox{reported} by the Particle Data 
Group are accommodated by Rarita--Schwinger (RS) type representations
$\lbrace{k\over 2},{k\over 2}\rbrace  \otimes \lbrack\lbrace
{1\over 2},0\rbrace \oplus\lbrace 0,{1\over2}\rbrace\rbrack $ with 
$k=1,3$, and 5, the highest spin states being $J=3/2^-$, $7/2^+$, and 
$11/2^+$, respectively. In contrast to this, most of the ``missing'' 
resonances fall into the opposite parity RS fields of highest-spins 
$5/2^-$, $5/2^+$, and $9/2^+$, respectively. Rarita--Schwinger fields 
with physical resonances as lower-spin components can be treated as a 
whole without imposing auxiliary conditions on them. Such fields do not
suffer the Velo--Zwanziger problem but propagate causally in the
presence of electromagnetic fields. The pathologies associated with RS
fields arise basically because of the attempt to use them to describe
isolated spin-$J=k+{1\over 2}$ states, rather than multispin-parity
clusters. The positions of the observed RS clusters and their spacing
are well explained trough the interplay between the rotational-like
${k\over 2}\left({k\over 2}+1\right)$-rule and a Balmer-like ${{-1}
\over {(k+1)^2}}$-behavior.
\end{abstract}

\setcounter{footnote}{0}
\renewcommand{\thefootnote}{\alph{footnote}}
\vspace*{0.6cm}
\input{psfig.sty}

\sloppy

\section{Baryons as Multispin-Parity Clusters}
\subsection{Introductory remarks}
One of the basic quality tests for any model of composite systems is
the level of \hbox{accuracy} reached in describing the excitation 
spectrum of its simplest object. \hbox{Recall}, that quantum mechanics 
was established only after the successful description of the spectrum of 
the hydrogen atom. Also one of the main successes of quantum 
electro\-dynamics as a gauge theory, the explanation of the Lamb shift 
of the electron levels in the hydrogen atom, was related to a 
spectroscopic observable. The history of spectroscopy is basically the 
history of finding the degeneracy symmetry of the spectra 
(see Ref.~\cite{Sternberg} for a related historical review). It is 
therefore, natural, to expect from the models of hadron structure that 
they should supply us with a a satisfactory description of baryon 
excitations. In that respect, the knowledge on the degeneracy group of 
baryon spectra appears as a key tool in constructing the underlying strong 
interaction dynamics. To uncover it, one has first to analyze isospin by 
isospin how the masses of the resonances from the full listing in 
Ref.~\cite{Part} spread with spin and parity. Such an analysis has been 
performed in our previous work,\cite{Ki97-98a} where it was found that 
Breit-Wigner masses reveal on the $M/J$ plane a well pronounced spin- and 
parity clustering. There, it was further shown that the quantum numbers of
the resonances belonging to a particular cluster fit into Lorentz
group representations of the type $\big\{{k\over 2}, {k\over 2}\big\}
\otimes \big[\big\{{1\over 2},0\big\} \oplus \big\{0,{1\over
2}\big\}\big]$, known as Rarita--Schwinger (RS) fields. To be
specific, one finds the three RS clusters with $k=1,3$, and $5$ in
both the nucleon $(N)$ and $\Delta$ spectra. These representations
accommodate states with different spins and parities and will be
referred to as multispin-parity clusters. To illustrate this
statement it is useful to recall that the irreducible representations
(irreps) $\big\{{k\over 2},{k\over 2}\big\}$ of the Lorentz group
yield in its Wick rotated compact form O(4), four-dimensional
ultraspherical harmonics, here denoted by $\sigma_\eta$, with
$\sigma=k+1$, where $\sigma$ is related to the principal quantum
number of the Coulomb problem, while $\eta $ is related to
\hbox{parity.} These O(4) irreps will be occasionally referred to in the
following as {\it Coulomb multiplets\/}. The latter are known to
collect mass degenerate O(3) states of integer internal angular
momenta, $l$, with $l=0,\dots,\sigma-1$. All three-dimensional
spherical tensors (denoted by $\sigma_{\eta;lm}$) participating the
ultraspherical one, have either \hbox{natural} ($\eta =+$), or unnatural
($\eta = -$) parities. In other words, they transform with respect to
the space inversion operation ${\cal P}$ as
\begin{eqnarray}	
{\cal P} \sigma _{\eta ;lm} = \eta e^{i\pi l}\sigma_{\eta;l-m}\,, \quad
l=0^\eta,1^{-\eta},\dots,(\sigma -1)^{-\eta}\, , \quad 
m=-l,\dots,l\, .
\label{party}
\end{eqnarray}

\noindent
In coupling a Dirac spinor, $\big\{{1\over 2},0\big\}
\otimes\big\{0,{1\over 2}\big\} $, to the Coulomb multiplets
$\big\{{k\over 2},{k\over 2}\big\}$ from above, the spin ($J$) and
parity $(P$) quantum numbers of the baryon resonances are created as
\begin{eqnarray}	
J^P = {1\over 2}^\eta,{1\over 2}^{-\eta}, {3\over 2}^{-\eta}, \dots,
\bigg(k+{1\over 2}\bigg)^{-\eta} \, , \quad k=\sigma -1\, .
\label{coupl_scheme}
\end{eqnarray}
The RS fields are finite-dimensional nonunitary representations of the
Lorentz group which, in being described by {\bf totally symmetric}
traceless rank-$k$ Lorentz tensors with Dirac spinor components,
$\psi_{\mu_1\mu_2\cdots \mu_k}$, have the appealing property that
spinorial and four-vector indices are separated. They satisfy the
Dirac equation \hbox{according~to}:
\begin{eqnarray}	
(i\partial\cdot \gamma - M)\Psi_{\mu_1 \mu_2\cdots \mu_k} = 0\, .
\label{Dirac_Proca}
\end{eqnarray}
{In terms of the notations introduced above, all reported baryons with
masses \hbox{below} 2500~MeV, are completely accommodated by the RS fields
$\psi_\mu $, $\psi_{\mu_1\mu_2\mu_3}$, and $\psi_{\mu_1\mu_2\cdots
\mu_5}$, having states of highest spin-$3/2^-$, $7/2^+$, and $11/2^+$,
respectively (see Figs.~1 and 2). In each one of the nucleon,
$\Delta$, and $\Lambda$ hyperon spectra, the RS cluster of lowest mass
is always $\psi_\mu $. For the nonstrange baryons, the $\psi_\mu$
cluster is followed\break}
\pagebreak

\begin{figure}[htbp]				
\centerline{\psfig{figure=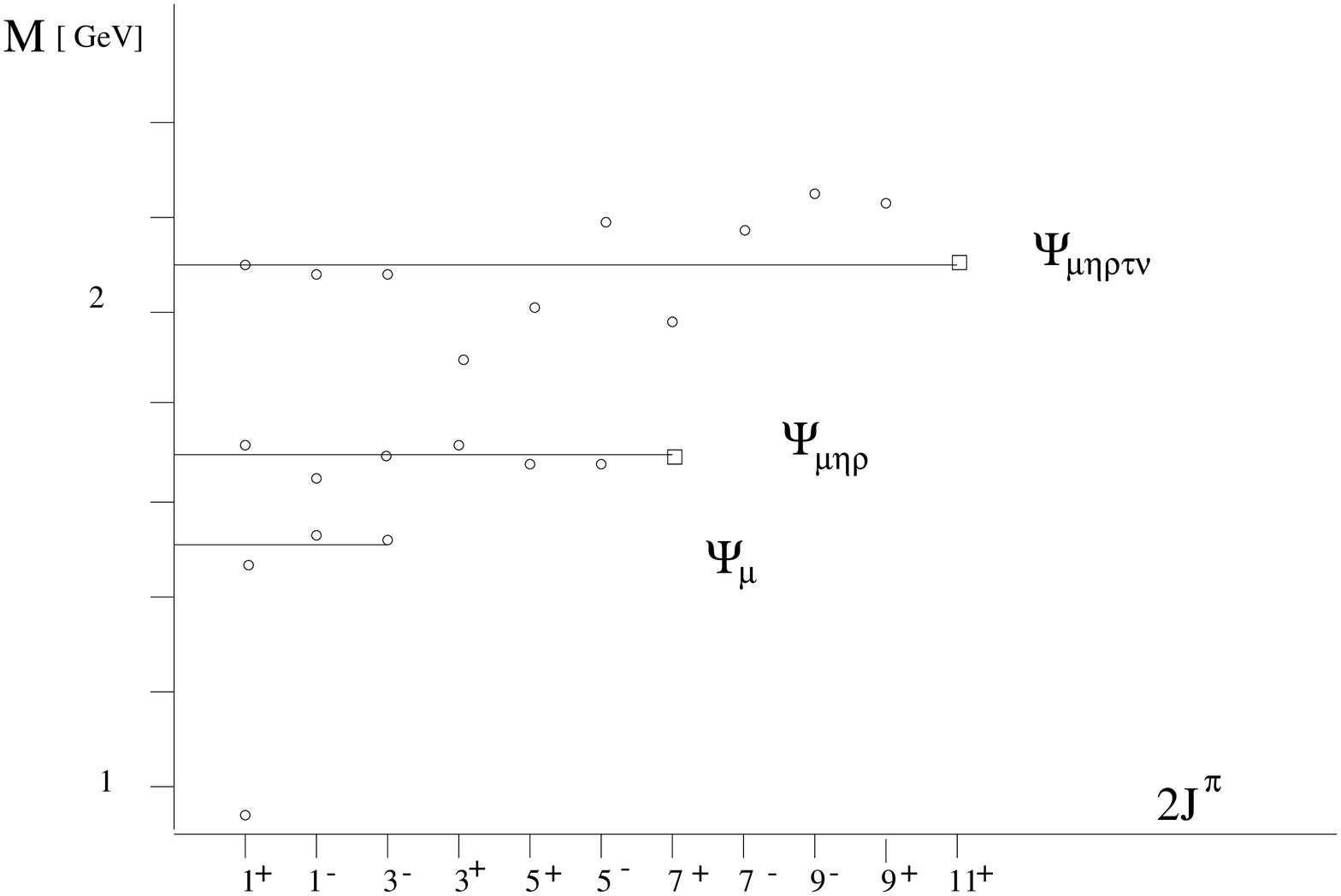,width=10cm}}
\vspace{0.2cm}
{\small The clustering of the reported nucleon excitations in terms
of the Rarita--Schwinger multispinors. Here, $\psi_\mu$ describes the
first $P_{11}$, $S_{11}$, and $D_{13}$ resonances. The second
$P_{11}$, $S_{11}$, and $D_{13}$ states together with the first
$P_{13}$, $D_{15}$, $F_{15}$, and $F_{17}$ states belong to
$\psi_{\mu\eta\rho}$. Finally, the third $S_{11}$, $P_{11}$, and
$D_{13}$ resonances, the second $P_{13}$, $D_{15}$, $F_{15}$, and
$F_{17}$ states, together with the first $G_{17}$, $G_{19}$, $H_{1 9}$
and $H_{1,11}$ states belong to $\psi_{\mu\eta\rho\tau\nu}$. The two
``missing'' $F_{17}$ and $H_{1,11}$ resonances have been denoted by
the open square $\Box $.
}
\end{figure}
\vspace*{30pt}

\begin{figure}[htbp]				
\centerline{\psfig{figure=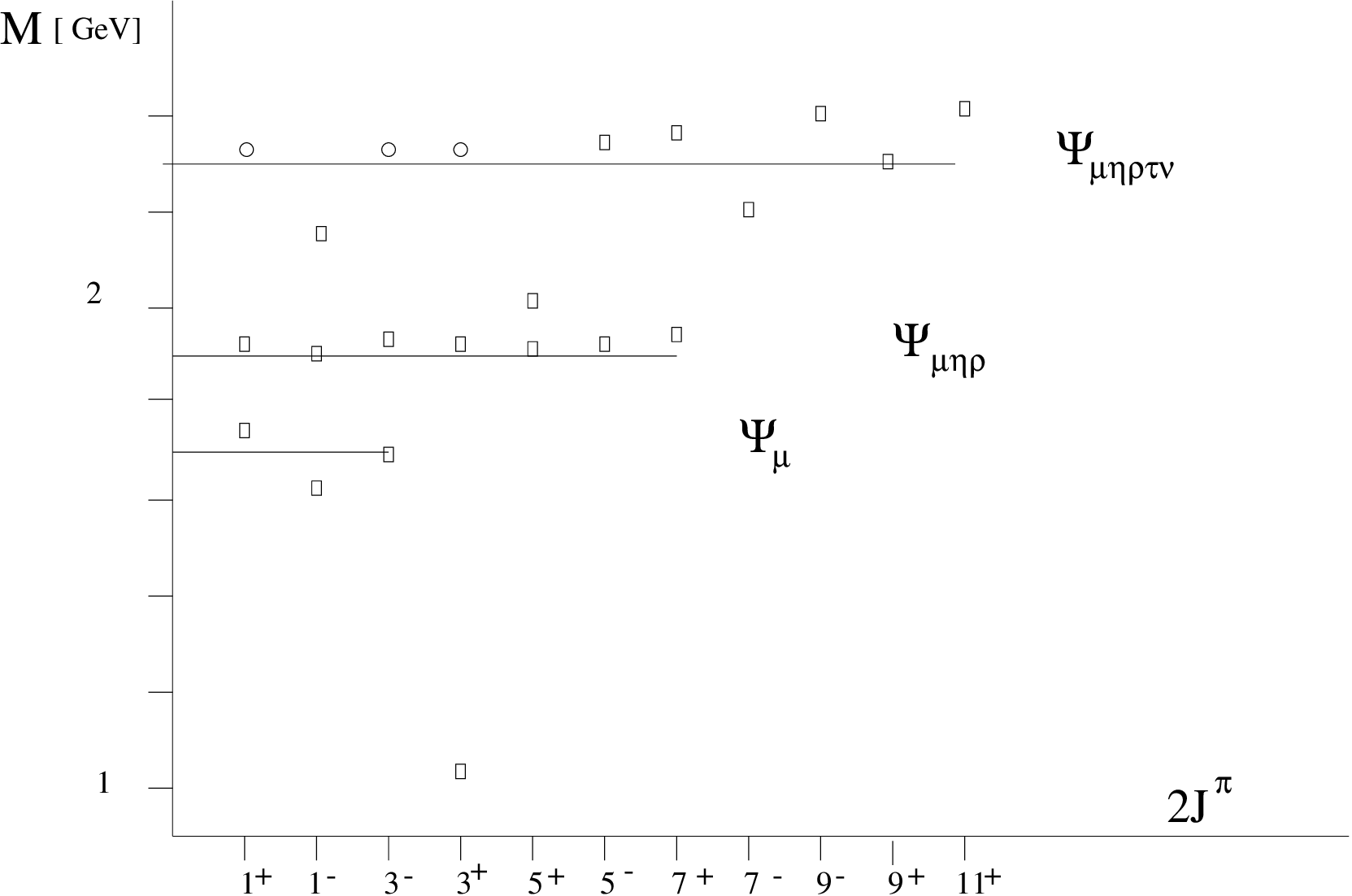,width=10truecm}}
\vspace{0.2cm}
{\small The clustering of the reported $\Delta$ excitations in terms
of the Rarita--Schwinger multispinors. Despite of its low mass, the
two-star $F_{37}$(2000) resonance fits well into the third RS cluster.
The three ``missing'' $P_{31}$ and $P_{3 3}$, and $D_{33}$ resonances
from the 2nd cluster have been marked by the open circle
$\circ$. Other notations as in Fig.~1.}
\end{figure}

\noindent
by $\psi_{\mu_1\mu_2\mu_3}$, and
$\psi_{\mu_1\mu_2\cdots \mu_5}$, while for the $\Lambda$ hyperons a
parity doubling of the resonances starts above 1800~MeV. In the
following we will extend the notation of the RS clusters to include
isospin $(I)$ according to $ \sigma_{2I,\eta}$. For example, the first
nucleon cluster is denoted by $2_{1,+}$, while $2_{3,+}$ and $2_{0,+}$
stand in turn for the corresponding $\Delta $- and $\Lambda $-hyperon
ones. From Eqs.~(\ref{party}) and (\ref{coupl_scheme}) follows that
the $2_{2I,+}$ clusters, where $I=1/2, 3/2$, and $0$, always unite the
first spin-${1\over 2}^+$, ${1\over 2}^-$, and ${3\over 2}^-$
resonances.

Indeed, the relative $\pi N$ momentum $L$ takes for $l=0^+$ the value
$L=1^+$ and corresponds to the $P_{2I, 1}$ state, while for $l=1^-$ it
takes the two values $L=0^-$, and $L=2^-$ describing in turn the
$S_{2I, 1}$ and $D_{2I, 3}$ resonances. The natural parity of the
first O(4) harmonics reflects the arbitrary selection of a scalar
vacuum through the spontaneous breaking of chiral symmetry. Therefore,
up to the three lowest $N$, $\Delta$, and $\Lambda$ excitations,
chiral symmetry is still in the Nambu--Goldstone mode. The Fock space
of the $2_{2I,+}$ clusters will be denoted in the following by ${\cal
F}_+$. Note, that in this context the first $P_{11}$ and $S_{11}$
states do not pair, because their internal orbital angular momenta
differ by one unit, instead of being equal but of opposite parities.
All the remaining nonstrange baryon resonances have been shown in
Ref.~\cite{Ki97-98a} to belong to either $4_{2I, -}$, or
$6_{2I,-}$. They have been viewed to reside in a different Fock
space, ${\cal F}_-$, which is built on top of a pseudoscalar vacuum.
For example, one finds all the seven $\Delta$-baryon resonances
$S_{31}$, $P_{31}$, $P_{33}$, $D_{33}$, $D_{35}$, $F_{35}$ and
$F_{37}$ from the $4_{3,-}$ cluster to be squeezed within the narrow
mass region from 1900~MeV to 1950~MeV, while the $I=1/2$ resonances
paralleling them, of which only the $F_{17}$ state is still
``missing'' from the data, are located around
1700$^{+20}_{-50}$~MeV. Therefore, the F$_{17}$ resonance is the only
nonstrange state with a mass below 2000 MeV which is ``missing'' in
the present RS classification scheme (compare Figs.~1 and 2).

In continuing by paralleling baryons from the third nucleon and
$\Delta$ clusters with $\sigma=6$, one finds in addition the four
states $H_{1, 11}$, $P_{31}$, $P_{33}$, and $D_{33}$ with masses above
2000~MeV to be ``missing'' for the completeness of the new
classification scheme. The $H_{1, 11}$ state is needed to parallel the
well established $H_{3, 11}$ baryon, while the $\Delta$-states
$P_{31}$, $P_{33}$, and $D_{33}$ are required as partners to the (less
established) $P_{11}$(2100), $P_{13}$(1900), and $D_{13}$(2080)
nucleon resonances.

The degeneracy group of the nonstrange baryon spectra found in
Refs.~\cite{Ki97-98a} on the grounds of the successful RS
classification is
\begin{equation}	
SU(2)_I\otimes SU(3)_c\otimes O(1,3)_{ls}\, .
\label{Lor_sp}
\end{equation}

Within this scheme, the approximately equidistant cluster spacing of
about 200 to 300~MeV between the mass centers of the RS clusters
appearing now, is by a factor 3 to 6 larger as compared, for example,
to the maximal mass splitting of 50--70~MeV within the $2_{1,+}$,
$2_{3,+}$, $4_{1,-}$, and $4_{3,-}$ multiplets.

\subsection{Multispin-parity clusters versus 
$SU(6)_{SF}\otimes O(3)_L$ multiplets} 

Traditionally, hadrons are classified in terms of 
$SU(6)_{ SF}\otimes O(3)_L$ multiplets. This classification is 
one of the most important paradigms in hadron spectroscopy. 
\hbox{According} to it, states like, say, the positive parity resonances 
$P_{13}$(1720), $F_{15}$(1680), $F_{35}$(1905), and $F_{37}$(1950), are 
viewed to belong to a 56(2$^+)$-plet, the $P_{11}$(1710) excitation is 
treated as a member of a $70(0^+)$-plet, while the negative parity baryons 
$S_{11}$(1535), $D_{13}$(1520), $S_{11}$(1650), $D_{13}$(1700), and 
$D_{15}$(1675) are assigned to a 70(1$^-$)-plet. The above examples 
clearly illustrate that states from our RS clusters separated by only few 
MeV, such as $D_{15}$(1675), $F_{15}$(1680), and $P_{11}$(1710) from 
$4_{1-}$, are distributed over three different 
$SU(6)_{\rm SF}\otimes O(3)_L$ representations, while on the other hand, 
resonances from different RS ``packages'' separated by about 200~MeV, 
such as $D_{13}$(1520) from $2_{1+}$, and $D_{13}$(1700) from $4_{1-}$, 
are assigned to the same multiplet.\cite{Part,Bhaduri} 
This means that $SU(6)_{ SF}\otimes O(3)_L$ can not be viewed as the 
degeneracy group of baryon spectra. Further, the 
$SU(6)_{ SF}\otimes O(3)_L$ multiplets appear approximately only 
``half-filled'' by the reported resonances. Several dozens states 
are ``missing'' for the completeness of this classification scheme 
(see Refs.~\cite{Part} and \cite{Bhaduri}for reviews). As long as 
observed and ``missing'' states are part of same multiplets, they are 
indistinguishable from the viewpoint of the underlying 
$SU(6)_{ SF}\otimes O(3)_L$ symmetry and there is no reason not to 
believe to the observability of all states. On the contrary, within the 
RS scheme, observed and ``missing'' states will fall apart and be 
attributed to Lorentz multiplets of different space--time properties. 
Therefore, they will become {\it distinguishable\/}. Within such a 
scenario, reasons for a possible suppression of states can be sought for.

The considerations from above show that the nonstrange baryons do not
exist as isolated higher-spin states but rather as multispin parity
clusters described in terms of particular RS representations of the
Lorentz group.

\section{Rarita-Schwinger Fields without Auxiliary Conditions}
Originally, the Rarita-Schwinger fields acquired importance in
connection to one of the longstanding problems in baryon
spectroscopy-- the relativistic description of higher-spin
states. Since Weinberg's work on particles with any spin\cite{J00J} it
is of common use to describe spin-$J$ baryons as the highest-spin
states of the Rarita--Schwinger representations of the Lorentz group
from Eq.~(\ref{Dirac_Proca}). In this way, the spin-$J=k+{1\over 2}$
field appears as the highest-spin state within the multispin-parity
cluster and in order to describe it, one has to eliminate all the
lower-spin components by suitably chosen auxiliary conditions. In
order words, one considers the lower-spin components as redundant,
unphysical states. The auxiliary conditions of common use read:
\begin{eqnarray}	
\bigg(g^{\nu \mu_1} + {1\over M^2}\partial^\nu \partial^{\mu_1}\bigg)
\Psi_{\mu_1 \mu_2\cdots \mu_k} &=& 
\Psi^\nu_{\,\,\,\mu_2\cdots\mu_k}\,, \\[5pt]
\gamma^{\mu_1}\psi_{\mu_1 \mu_2 \cdots \mu_k} &=& 0\, .
\label{redundand_comp}
\end{eqnarray}
The first auxiliary condition is nothing but the Proca equation for a
four vector. It is equivalent to $\partial^{\mu_1}\psi_{\mu_1 \mu_2
\cdots \mu_k}=0$ and eliminates  the time component of the Lorentz
vector (pseudovector) associated with the index $\mu_1$, while the
second condition \hbox{excludes} its longitudinal degree of freedom. 
As a result, the spinor couples to the purely transversal components 
of the $\mu_1$ vector, and gives rise to a spin-3/2$^\pm$ field. In
repeating same procedure $k$ times, the highest-spin state with
$J=k+{1\over 2}$ is created. The auxiliary conditions
prevent, therefore, that the minimal helicities $\pm 1/2$ of the
spin-3/2 field get mixed up with such corresponding to the
spin-$1/2^+$, and $1/2^-$ members of $\psi_{\mu_1}$,
respectively.\footnote{As long as the multispinor-tensor is totally
symmetric with respect to the Lorentz indices, it was sufficient to
impose the auxiliary conditions to one index only.} It is well known
that the procedure of the exclusion of the lower-spin component from
the RS multispin-parity clusters suffers several drawbacks. For
example, when minimally coupled to an external electromagnetic field,
the propagation of the truncated RS field may violate causality. This
drawback is known as the Velo--Zwanziger problem.\cite{EriceZw}
Furthermore, in trying to incorporate the two auxiliary conditions
into the Lagrangian of the RS fields, one is forced to introduce an
arbitrary parameter to account for the fact that for off-mass shell
particles, the auxiliary conditions are not any longer valid and the
exclusion of the lower-spin components not any longer guaranteed.
Several solutions to this problem have been suggested over the
decades, the only reliable being the concept of supersymmetry.

Beyond the RS formalism, other methods for describing higher-spin
states have been explored in the literature. In Ref.~\cite{DVA1}
causal higher-spin propagators have been constructed by Ahluwalia {\it
et~al.\/} for $(J,0)\oplus (0,J)$ states and shown to be necessarily
nonlinear in the momenta due to the $p^{2J}$-dependence of the boost
in such representation spaces. This is a promising method, though the
techniques for calculating three point functions including besides the
spin-$J$ state, a nucleon and an external vector field, need to be
still developed, in particular because the $(J,0)\oplus (0,J)$ fields
do not carry spinor-vector indices. Those states may acquire
significant importance in future research on the CPT structure of
space--time, especially because their $C$ and $P$ properties turn out
to be quite different from those of the RS fields.\cite{DVA2}

More recently, higher-spin fermion states have been constructed by
Moshinsky {\it et~al.\/} as eigenstates of the relativistic Dirac
oscillator (see Ref.~\cite{Moshinsky} for a review). In exploiting
the circumstance, that the elements of the Clifford algebra of the
Dirac matrices generate the group O(6), and that the Hamiltonian
contains only some of the generators of its subgroup O(5), the wave
function of a spin-$J$ state has been expanded into the basis of O(5)
irreps by means of a properly constructed Racah algebra that accounts
for the reduction of the O(5) down to O(3) through the chain
O(5)/O(4)/O(3). There, the RS fields appear as intermediate and their
lower-spin components are automatically projected out by the Racah
algebra. This scheme, which has been developed basically for the needs
of the relativistic description of many body systems, such like
nuclei, has the immediate advantage of being straightforward in
calculating various transition matrix elements. The properties of the
relativistic harmonic oscillator have been extensively studied by Kim
and Noz (see Ref.~\cite{KiNo} for a textbook
presentation). Applying these ideas to baryon clustering is a task
worthy to be explored.

On the other side, the RS spinors {\it without\/} auxiliary conditions
have the great advantage that they do not suffer the pathologies
indicated above. As we already argued in the previous section, exactly
this very case is relevant for baryon spectroscopy. In view of this
relevance we here consider the properties of the simplest RS field
$\psi_\mu $ without imposing any subsidiary conditions on it. As
noticed by Kruglov and Strashev in Ref.~\cite{Kruglov86}, the Dirac
equation $(p^\mu \gamma_\mu-M)u_\nu ({\bf p})=0$ for the vector-spinor
is equivalent to
\begin{equation}	
(p^\mu\Gamma_\mu - M)U({\bf p}) = 0\, ,
\label{Krug}
\end{equation}
where $\Gamma_\mu =1\!\!1_4\otimes \gamma_\mu$, and $U({\bf p})$ is a
16-component vector. The conjugate quantity $\bar U({\bf p}) $ is
defined as $U^\dagger ({\bf p})\eta $ with $\eta = \Pi\otimes
\gamma_0$ and $\Pi$=diag$(1,-1,-1,-1)$. It can
be proven that the matrices $\Gamma_\mu$ replicate in the
16-dimensional space the anticommutation relations of the Dirac
matrices. This allows one to gauge the corresponding Lagrangian,
${\cal L}=\bar U p^\mu\Gamma_\mu U -M\bar U U$ in the usual way by
introducing the covariant derivative as $D_\mu = \partial_\mu -ie
A_\mu $. The propagator $S$ of the $U$ field without auxiliary
conditions constructed in this way reads:
\begin{equation}	
S = {{p^\mu\Gamma_\mu +M}\over {2M (p^2-M^2)}}\, .
\end{equation}
It is causal and does not create any problems. It is important to
note that also the cluster propagator $S_{\mu\nu}$ for the $u_\mu
({\bf p})$-field with Proca's auxiliary condition,
\begin{equation}	
S_{\mu \nu} ={{\Big(g_{\mu\nu}- {{p^\mu p^\nu}\over M^2}\Big)
(p\!\!/+M)} \over {2M (p^2-M)}}\, ,
\end{equation}
is causal as both the Dirac and Proca equations do not change form
after gauging. It is the second auxiliary condition in
Eq.~(\ref{redundand_comp}) that is responsible for the occurrence of
the Velo--Zwanziger problem.

To construct a $N\gamma\psi_\mu $ vertex one may notice that the
nucleon--photon system $N\gamma$ carries same indices like $\psi_\mu$
since the photon transforms as a Lorentz vector $V_\mu$, while the
nucleon $(N)$ is a Dirac spinor. Therefore, the $N\gamma$ system can
be converted similarly to Eq.~(\ref{Krug}) to a 16-dimensional spinor.
Same is true for the $N\pi\psi_\mu $ vertex, provided the pion is
gradiently coupled, so that $(\partial^\mu\pi)N$ transforms as a
Lorentz pseudovector, $A_\mu$, with Dirac spinor components. With
that the description of multispin-parity clusters as intermediate
states in processes such like $\pi (\eta)$ photo production off
proton is straightforward.

In the previous section we showed that exactly this is the case in
baryon spectra. There, we showed, that in the excitation spectra of
the nonstrange baryons one does not find isolated higher-spin states
but rather complete multispin-parity clusters of the Rarita--Schwinger
type. The result followed from the analysis of the degeneracy group
in the baryon spectra. In the next section we are going to study the
RS clustering of baryons on the level of the quark degrees of freedom.

\section{Decomposing Three Quark Hilbert Space into Lorentz Group
Representations} To get a deeper insight into the structure of baryon
spectra, it is quite instructive to consider as an illustrative
example how the quantum numbers of the baryon excitations can emerge
from a configuration space with one excited quark only. We begin with
the simplest space spanned by the $1s$, $1p$, and $2s$-single-particle
shells. We use the familiar shell-model notation of the
single-particle basis, where $1s$ means the first shell with zero
orbital angular momentum (a.m.), $1p$ is the first shell with a unit
a.m. etc.\footnote{The configurations considered can be interpreted as
leading components of the wave functions of baryons viewed as
three-quark states. The debates about possible non-$qqq$ character of
resonances will be ignored so far.} The one-quark excitations give
rise to the following orbitally excited two-quark configurations (in
standard shell-model notations) of both natural and unnatural
parities:
\begin{equation}	
\Big[1s_{{1\over 2}}^{1}\otimes 2s_{{1\over 2}}^1\Big]^{l=0^+,1^+}\, ,
\quad \Big[1s_{{1\over 2}}^{1} 
\otimes 1p_{{1\over 2}}^1\Big]^{l=0^-,1^-}\, ,
\quad \Big[1s_{{1\over 2}}^{1} 
\otimes 1p_{{3\over 2}}^1\Big]^{l=1^-, 2^-}\,.
\label{1P_1H}
\end{equation}
Note that we here consider the quark model in the $j$--$j$ rather that
in the LS coupling ordinarily exploited in the traditional
$SU(6)_{SF}\otimes O(3)_L$ quark models. Note, further, that the
spin-$1^-$ state appears twice in the latter equation as it can result
from both the $\Big[1s_{{1\over 2}}^{1}\otimes 1p_{{1\over
2}}^1\Big]$, and $\Big[1s_{{1\over 2}}^{1}\otimes 1p_{{3\over
2}}^1\Big]$ configurations. The quantum numbers ($0^+, 1^-$) fit into
the $\big\{{1\over 2}, {1\over 2}\big\}_+ $ polar Lorentz vector,
denoted by $V_\mu$
\begin{equation}	
V_\mu = \bigg\{{1\over 2}, {1\over 2}\bigg\} _+: \ 
\Big(\Big[1s^{1}_{{1\over 2}} \otimes 2s^1_{{1\over 2}}\Big]^{l=0^+}\, ,
\quad \Big[1s^{1}_{{1\over 2}} 
\otimes 1p^1_{{1\over 2}}\Big]^{l=1^-}\,
\Big) .
\label{natural_par}
\end{equation}
In the following, the index $+/- $ attached to a Lorentz group
representation will be used to label natural/unnatural parities. The
spin-parity sequence ($0^-, 1^+,2^-$) fits into $\{1,1\} _-$, the
totally symmetric 2nd rank Lorentz pseudotensor $A_{\mu_1\mu_2}$
according~to
\begin{eqnarray}	
A_{\mu_1\mu_2} &=& \{1, 1\} _- : \ 
\Big[1s^{1}_{{1\over 2}}\otimes 1p^1_{{1\over 2}}\Big]^{l=0^-}\, , 
\quad \Big[1s^{1}_{{1\over 2}}\otimes 2s^1_{{1\over 2}}\Big]^{l=1^+}\, ,
\quad \Big[1s^{1}_{{1\over 2}}\otimes 
1p^1_{{3\over 2}}\Big]^{l=2^-}\, .  
\label{unnat_par}
\end{eqnarray}
Finally, the remaining natural parity spin-$1^-$ state from 
$\Big[1s_{{1\over 2}}^1\ 1p_{{3\over 2}}^1\Big]$ can be viewed as
$\{1,0\}_+$ and it arises out of the multiplicity of the spin-$1^-$
two-quark state in Eq.~(\ref{1P_1H}). As a result, the Hilbert space
under consideration, (it will be denoted by ${\cal
H}^{2q}_{1s-2s-1p}$) decomposes into Lorentz group representations as
follows:
\begin{equation}	
{\cal H}^{2q}_{1s-2s-1p}=\bigg\{{1\over 2}, {1\over 2}\bigg\}_+ \oplus
\{1,1\} _- \oplus \{1,0 \}_+ \, .
\label{H_decomp}
\end{equation}
In other words, ${\cal H}^{2q}_{1s-2s-1p}$ is decomposed into an odd
spin-$1^-$ state and the pair of representations $A_{\mu_1 \mu_2}$,
and $V_\mu$, describing in turn states of unnatural and natural
parities:
\begin{equation}	
{\cal H}^{2q}_{1s-2s-1p}= V_\mu \oplus A_{\mu_1\mu_2} 
\oplus \{1,0\}_+ \, ,
\label{H1_Tdec}
\end{equation}

In noting that ${\cal H}^{2q}_{1s-2s-1p}$ corresponds to $n=2$, one
finds as quite an interesting result that the principal quantum number
$n$ of the Coulomb problem associated with the single particle shells
in the $2q$-Hilbert space gives rise to ultraspherical O(4) harmonics
with $\sigma=n$, and $\sigma=n+1$, respectively. While the $\sigma=n$
ultraspherical harmonics contain O(3) states of natural parities,
those with $\sigma=n+1$ contain unnatural parities states.

A similar analysis of the $3s-2p-1d$ configuration (it corresponds
to a principal quantum number $n=3$ of the Coulomb problem) shows that
${\cal H}_{1s-3s-2p-1d}^{2q}$ decomposes into
\begin{equation}	
{\cal H}^{2q}_{1s-3s-2p-1d} = 
\{1,1\}_+\oplus \bigg\{{3\over 2}, {3\over 2}\bigg\}_-
\oplus [\{1,0\} \oplus \{0, 1\}] \oplus \{2,0\}_+ \, .
\label{n_3}
\end{equation}
In terms of Lorentz-vector index notations one finds
\begin{equation}	
{\cal H}^{2q}_{1s-3s-2p-1d} = 
V_{\nu_1\nu_2} \oplus A_{\mu_1\mu_2\mu_3}\oplus 
F_{\mu \nu} \oplus \{2,0\}_+\, .
\label{h_3}
\end{equation}
Here, the $\{1,0\}\oplus\{0,1\}$ representation has been described in
terms of the totally antisymmetric 2nd rank Lorentz tensor
$F_{\mu\nu}$ rather than as a six component vector. 
{}Finally, in including the high-lying $1f$ and $1g$ shells, one finds
\begin{eqnarray}	
{\cal H}^{2q}_{1s-4s-3p-2d-1f-1g} = V_{\mu_1...\mu_4} \oplus 
A_{\nu_1\cdots \nu_5} \oplus \sum_{m=1}^{m=n-2}[\{m,0\} \oplus\{0, m\}]
\oplus \{4,0\} _+\, .
\label{h_5}
\end{eqnarray}
The extension of the $4s-3p-2d-1f$ single-particle space to include
the $1g$ shell is supported by the empirical observations. In doing
so, we effectively produced the quantum numbers of orbital angular
momenta $l=0,\dots ,4$ which correspond to $n^{\rm eff}=5$. The
latter notation has been introduced in order to distinguish $n^{\rm eff}$
from the genuine principal quantum number $n=5$ of the Coulomb problem
associated with the $5s-4p-3d-2f-1g$ space.

Equation~(\ref{h_5}) shows than any ${\cal H}^{2q}_n$ contains fields
of natural, ($V_{\mu_1\mu_2\cdots \mu_{n-1}}$), and unnatural
($A_{\mu_1\mu_2\cdots \mu_n}$) parities, respectively. They are
supplemented by several parity symmetric $\{m,0\}\oplus\{0,m\}$ with
$m=1,\dots, n-2$ states, as well as by an odd $\{n-1,0\}_+$ state.

Representations of the type $\{m,0\} \oplus \{0,m\}$ deserve special
attention. To be specific, the $\{1,0\} \oplus \{0,1\} $ state, when
considered as a totally symmetric (Sym) second order tensor
$f^\alpha_\beta $ with $SL(2,C)$ spinor ($\xi^\alpha)$ components
i.e. as $f^\alpha_\beta =$Sym$\xi^\alpha\xi_\beta$, describes boson and
antiboson of same parities (see Ref.~\cite{KiNo} for an
understandable and yet precisely written textbook on Lorentz group
representations). Also the selfconjugate $\big\{{1\over 2}, {1\over
2}\big\}$ representation predicts this usual type of bosons. In
contrast to this, it was shown in Ref.~\cite{DVA2} that the
$\{1,0\} \oplus \{ 0,1\}$ representation can accommodate an entirely
new type of boson whose parity is opposite to its antiparticle. This
was achieved by completely giving up the $SL(2,C)$ tensor--spinor
technique for constructing irreducible representations of the Lorentz
group. Instead, spin-1 bosons and antibosons were described in terms
of a set of fundamental six component $U$-, and $V$ spinors,
respectively. In explicitly constructing then the charge $(C)$ and
parity $(P)$ conjugation operators within this space as anticommuting
$6\times 6$ matrices, i.e. $\{C,P\}=0$, the parities of the $U$ and
$V$ spinors have been proven to be opposite. Such bosons have been
termed to as Bargman--Wigner--Wightman ones, or, shortly, BWW bosons,
a notation which we will adopt in the following. In this way it was
demonstrated in Ref.~\cite{DVA2} that the $\{1,0\} \oplus \{0,1\}$
representation, if treated beyond the $SL(2,C)$ spinor algebra
technique, allows for the embedding of entirely new type of bosons.
The above example illustrates that specification of the Casimir
invariants of the Poincar\'e group alone, i.e. mass and maximal spin
of the representation, do not determine in a unique way the relevant
Lorentz multiplet. Only invoking discrete symmetry transformations,
such as charge and parity conjugation, fixes the representation space
unambiguously (cf. Ref.~\cite{DVA3} for a discussion).

With this facts in mind, we now observe that the decompositions of
${\cal H}^{2q}_n$ contain together with representations describing
bosons of standard type, also such potentially capable of
describing  BWW bosons. Finally, the series always end
with an odd $\{n-1,0\}_+$ state of natural parity. The latter one can
be considered as the large component of a $\{n-1,0\} \oplus\{0, n-1\}$
$U$-spinor (in the notations of Ref.~\cite{DVA2}) and reflects the
inherent nonrelativistic element of the quarkish shell-model space.

Now one can couple the spectator $1s_{{1\over 2}}$ quark to the
two-quark configurations and obtain the quantum numbers of the
baryons. To be specific, we will carry this procedure for the case of
Eq.~(\ref{1P_1H}). In doing so, one ends up with a spin-parity
sequence containing the ten states
\begin{equation}	
{1\over 2}^+; \ {1\over 2}^+,{3\over 2}^+; \ 
{1\over 2}^-; \ {1\over 2}^-,{3\over 2}^-; \ 
{1\over 2}^-, {3\over 2}^-; \ {3\over 2}^-,{5\over 2}^-\, .
\label{full_space}
\end{equation}
They can be distributed over three different Lorentz representations.
The latter are obtained from coupling the Dirac spinor
$\big\{{1\over2},0\big\} \oplus \big\{0,{1\over 2}\big\}$, to be
denoted by $\psi $ in the following, to the Lorentz tensors in
Eq.~(\ref{H_decomp}). In doing so, one first finds
\begin{eqnarray}	
V_\mu \otimes \psi &=:& \Big(\Big[1s^{-1}_{{1\over 2}} \otimes
2s^1_{{1\over 2}}\Big]^ {l=0^+} \otimes 
1s_{{1\over 2}}^1\Big)^{{1\over 2}^+} \, ,
\quad \Big(\Big[1s^{-1}_{{1\over 2}} \otimes 
1p^1_{{1\over 2}}\Big]^{l=1^-} \otimes 
1s_{{1\over 2}}^1\Big)^{{1\over 2}^-;{3\over 2}^-}\, .  
\label{natural_par2}
\end{eqnarray}
In terms of vector--spinor indices the last equation is converted to
\begin{equation}	
\psi_\mu := V_\mu \otimes\psi \, .
\label{RS_vector}
\end{equation}
Here, $\psi_\mu$ stands for a Lorentz vector with Dirac spinor
components. It describes a RS field with a spin-$3/2^-$ as the state
of highest spin. Therefore, at the baryon level, one finds $\psi_\mu
$ to include besides the highest spin-$3/2^-$ resonance, two more
spin-1/2 states carrying opposite parities.
\begin{eqnarray}	
\psi_\mu \to \bigg({1\over 2}^+; {1\over 2}^-, {3\over 2}^-\bigg)\, .
\label{21_np}
\end{eqnarray}

The remaining unnatural parity configurations can be accommodated by
the $\{1,1\}_- \otimes \psi $ RS representation. It has the unnatural
parity spin-$5/2^-$ as a \hbox{highest}~spin:
\begin{eqnarray}	
A_{\mu_1\mu_2}\otimes \psi &=:& \Big(\Big[1s^{-1}_{{1\over 2}}\otimes
1p^1_{{1\over 2};{3\over 2}}\Big]^{l=0^-, 2^-}\otimes
1s^1_{{1\over 2}}\Big)^{{1\over 2}^-; {3\over 2}^-, 
{5\over 2}^-}\, ,  \\[5pt]
&&{} \Big(\Big[1s^{-1}_{{1\over 2}}\otimes
2s^1_{{1\over 2}}\Big] ^{l=1^+}\otimes
1s^1_{{1\over 2}}\Big)^{{1\over 2}^+, {3\over 2}^+}\, .
\label{unnat_par2}
\end{eqnarray}
The RS field $\{1,1\}_-\otimes\psi$ from the last equation is
associated with the totally symmetric rank-2 Lorentz tensor with Dirac
spinor components $\psi_{\mu_1 \mu_2}$
\begin{eqnarray}	
\psi_{\mu_1\mu_2} := A_{\mu_1\mu_2} \otimes \psi\, .
\label{31_unn}
\end{eqnarray}
The $\psi_{\mu_1\mu_2}$ field includes, besides the highest
spin-$5/2^-$ state, also two more spin-1/2, as well as two more
spin-3/2 parity duplicated states according to
\begin{equation}	
\psi_{\mu_1\mu_2} \to\bigg({1\over 2}^-;{1\over 2}^+, {3\over 2}^+;
{3\over 2}^-, {5\over 2}^-\bigg)\, .
\label{5/2_RS}
\end{equation}
{}Finally, the two more natural parity states $\big({1\over 2}^-,
{3\over2}^-\big)$ are obtained from the coupling of the Dirac spinor to 
the ordinary three vector and are described by means of a three vector
with Dirac spinor components, $\psi_a$ as
\begin{eqnarray}	
\psi_a := V_a\otimes \psi\, , \quad a=1,2,3\, .
\end{eqnarray}
As a result, the $1s-2s-1p$ three-quark $(3q)$ Hilbert space contains
the two opposite parity RS clusters $\psi_\mu (x)$ and
$\psi_{\mu\nu}(x)$, respectively:
\begin{equation}	
{\cal H}^{3q}_{1s-2s-1p}\to [V_\mu \oplus A_{\mu_1\mu_2}\oplus V_a] 
\otimes \psi\, .
\label{H2_decomp}
\end{equation}
In proceeding in this way, all the quantum numbers of the excited
baryons resulting from a given Hilbert space can be obtained (see
Table~1).

\begin{table}[htbp]			
\caption{Decomposition of the three-quark Hilbert space
into Lorentz group representations. See text for notations.}
\vspace*{0.5cm}
{\begin{tabular}{@{}lccc@{}}
\hline\\[-6pt]
{resonance} & ${\cal H}_{1s-2s-1p}$ & ${\cal H}_{1s-3s-2p-1d}$
& ${\cal H}_{1s-4s-3p-2d-1f-1g}$ \\[3pt]
{status} & $n=2$ & $n=3$ & $n^{\rm eff}=5$ \\[3pt]
\hline\\[-6pt]
37 reported, & $V_\mu\otimes \psi$ & $A_{\mu_1\mu_2\mu_3}\otimes \psi$
& $A_{\mu_1\mu_2\mu_3\mu_4\mu_5}\otimes \psi$ \\[3pt]
5 ``missing'' & $(2_{2I,+})$ & $(4_{2I,-})$ & $(6_{2I,-})$ \\[3pt]
all ``missing'' & $A_{\mu_1\mu_2}\otimes\psi$ 
& $V_{\mu_1\mu_2}\otimes\psi$ & $V_{\mu_1\mu_2\mu_3\mu_4}
\otimes\psi$ \\[3pt] 
& $(3_{2I,-})$ & $(3_{2I,+})$ & $(5_{2I,+})$ \\[3pt]
all ``missing'' & $\{1,0\}_+\otimes\psi$ 
& $\{2,0\}_+\otimes\psi$ & $\{4,0\}_+\otimes\psi$ \\[3pt]
all ``missing'' & {} & $F_{\mu \nu}\otimes\psi$ 
& $\sum_{m=1}^{m=3}\{m,0\}\oplus\{0,m\} \otimes \psi$ \\[3pt]
\hline
\end{tabular}}
\end{table}

\section{O(4) Degeneracy in ${\cal H}^{3q}$ and Parity Classification of\\
``Missing'' Resonances}
In comparing now the quantum numbers of the resonances reported in the
full list by the Particle Data Group\cite{Part} to those contained in
${\cal H}^{3q}$ in Table~1, one immediately realizes that only a part
of the possible configurations has been detected so far. Namely, the
observed states fitted into the three RS representations from the
first row, from which only five resonances are still ``missed.''
These were the first $F_{17}$ and $H_{1,11}$ nucleon states, the third
$P_{31}$, and $D_{33}$, and the fourth $P_{33}$ $\Delta$ states. All
the remaining ``missing'' resonances belong to the representations
from the second row on in Table~1. With the exception of the $\Lambda$
hyperon $S_{01}$(1670) and $D_{03}$(1690) states, there are no other
resonances which can be interpreted as candidates for
$\{1,0\}_+\otimes\psi$ configurations.

In connection with this empirical observation, the question arises,
how to interpret the absence of the remaining resonances from Table~1
from the spectra. One possibility is, that they are suppressed by
some dynamical reasons, another one could be, that they have not yet
been observed. While the latter belief is the widely spread one, we
here are rather interested in exploring the former option.

One possible way towards explaining a suppression of the resonances
lying outside of the $2_{2I,+}$, $4_{2I,-}$, and $6_{2I,-}$ clusters
is to first introduce the O(4) degeneracy of the diquarks and then
consider the O(4) bosons (in fact, the Wick rotated O$(1,3)$ bosons)
created in this manner as fundamental point like degree of freedom
(d.o.f.). The most important O(4) bosons are the Coulomb multiplets
\begin{equation}	
{\cal C}_{\mu_1\mu_2\cdots \mu_k} (x)\simeq
\bigg\{{{k}\over 2}, {{k}\over 2}\bigg\} \to 
(0^\eta, 1^{-\eta}, \dots, k^{-\eta})\, ,
\label{hyper_quark}
\end{equation}
for which we coined the term ${\cal C}$-{\it hyperquark}.
In general, all bosons from Table 1 will be referred to as
hyperquarks. The ${\cal C}$ hyperquarks
behave with respect to space reflection as either Lorentz tensors, or, 
pseudotensors. In being fundamental point like subbaryonic degrees of 
freedom, they are created within a Fock space, ${\cal F}_{\eta}$, 
by operators, $C^\dagger_{\mu_1\mu_2\cdots \mu_k;lm}$, acting upon a 
vacuum, $|0^\eta \rangle$, of either positive ($\eta =+$), or
negative $(\eta = -$) parity. Here, $lm$ denote the quantum numbers
of the O(3) states accommodated by the O(4) harmonics. To be
specific, for $2_{1,+}$, or, $V_\mu\otimes \psi$ in equivalent
notation, one has
\begin{eqnarray}	
{\cal P} C_{\mu;lm}^\dagger|0^+\rangle = e^{i\pi l}
C^\dagger_{\mu ; l,-m}|0^+\rangle\, , \quad 
l=0,1\, , \quad m=-l,\dots, l\, .
\label{p-h_oper}
\end{eqnarray}

The idea of the point like character of the diquarks, and,
consequently, the hyper\-quarks, has been exploited in the literature to
reduce the three-quark Faddeev equations to a two-body quark--diquark
Bethe--Salpeter equation (see, for example Refs.~\cite{Hel97} and
\cite{Kus}). However, the last consequence of the point like nature
of a quantum object, the parity selection rule in Eq.~(\ref{p-h_oper})
due to its belonging to a Fock space with a vacuum of fixed parity,
was not considered by none of the diquark model versions. The
essential difference between the present quark--hyperquark model (QHM)
and the customary quark--diquark models (QDM) (see
Ref.~\cite{Anselmino} for a digest) is the \hbox{assumed} O(4) clustering
of the diquarks and the parity selection rule. To be specific, the
operator $D^\dagger_{2_1,+}$ which creates the lightest
${\cal C}$- hyperquark is defined as the following linear superposition:
\begin{eqnarray}	
\begin{array}{rcl}
D^\dagger_{2_1, +}|0^+ \rangle &=& \sum_{lm}\, 
c_{lm} C^\dagger _{\mu;lm}|0^+\rangle \, , \\[12pt]
\sum_{lm}|c_{lm}|^2 &=& 1\, , \\[16pt]
\langle 0^+| D^\dagger_{2_{1,+}}|0^+\rangle
&=& \sum_{lm}\, c_{lm}{\cal R}_{2 l} (r)
Y_{2 lm}(\alpha ,\theta ,\phi) \, .
\end{array}
\label{hyperquark}
\end{eqnarray}
In Eq.~(\ref{hyperquark}) the radial part of the ${\cal C}$-hyperquark 
wave function has been denoted by ${\cal R}_{\sigma l}(r)$, while its
angular part has been determined by the four-dimensional harmonics
$Y_{\sigma lm}$ defined in the standard way in Ref.~\cite{Shibuya}
as
\begin{eqnarray}	
Y_{\sigma lm}(\alpha ,\theta ,\phi) = i^{\sigma-1-l} 2^{l+1}l!
{{\sigma (\sigma -l-1)}\over {2\pi (\sigma +1)}}
\sin^\sigma \alpha\, {\cal C}^{l+1}_{\sigma -l-1}(\cos\alpha)
Y^l_m (\theta, \phi)\, .
\label{Gegenb}
\end{eqnarray}
Here, ${\cal C}^{l+1}_{\sigma -l-1} (\cos \alpha)$ denote the
Gegenbauer polynomials, while $Y^l_m(\theta ,\phi)$ are the standard
three-dimensional spherical harmonics.

In general, the Lorentz covariant isodoublet spin- and parity-cluster
$\sigma_{1,\eta}$, considered as a ${\cal C}$- hyperquark
coupled to a spectator $1s_{{1\over 2}}$ quark, is now described as
\begin{eqnarray}	
\sigma_{1, \eta} = \psi_\mu \otimes 
\big[T\otimes \chi^{{1\over 2}}\big]^{{1\over 2}}\, , \qquad
\psi_\mu = D^\dagger_{\sigma _{1,\eta}}b^\dagger_{1s_{{1\over 2}}}
|0^\eta \rangle\,, \label{island}
\end{eqnarray}
where  $\chi^{{1\over 2}} $ is the ordinary isospinor of
the spectator quark, while  $T$ stands for the isospin part of the 
hyperquark wave function. It should not be confused with the isospins 
attached to diquarks in the traditional quark-diquark models,
where the diquark isospin varies with $l$ and is 
determined from the requirement on a  totally antisymmetric 
3q-  wave function. Obviously, for all $\Delta $ states $T$ has to be
an isovector. In having once figured out the quantum numbers of the 
Lorentz multiplets under the guidance of Sect.~3, we here from now on 
design a model, where the hyperquarks are considered as fundamental 
degrees of freedom in their own rights. In such a case, 
antisymmetrization does not matter any longer, because the baryon 
constituents are essentially different particles.

The O(4) symmetry ansatz for the quark-hyperquark model assumed in the
present work is independently supported by the observed rapid
convergence of the covariant diquark models in the basis of the
Gegenbauer polynomials considered among others in Ref.~\cite{Hel97}
and \cite{Kus}. It is worthy of being pursued especially because of
the quite uncertain experimental status of the ``missing'' resonances.

Now, the absence of the $3_{2I.-}$ cluster associated with 
$A_{\mu_1\mu_2}\otimes \psi $ from Eq.~(\ref{unnat_par2}), may reflect
the circumstance, that the Nambu--Goldstone mode of chiral symmetry,
known to favor the natural parity for the ground state, extends to the
domain of the first $P_{2I,1}$, $S_{2I,1}$, and $D_{2I,3}$ resonances
from ${\cal H}_{1s-2s-1p}$. In this way the scale of the hidden mode
of that symmetry is predicted. The unnatural parity of the $4_{2I,-}$
and $6_{2I,-}$ clusters can be interpreted as a change of the vacuum
from scalar to pseudoscalar. Therefore, nucleon excitations to
resonances from the $4_{1,-}$ and $6_{1,-}$ clusters can be
interpreted as chiral phase transitions. The natural parity
$3_{2I,+}$, and $5_{2I+}$ clusters could be expunged from ${\cal F}_-$
by that very mode of chiral symmetry realization there. Finally, the
absence of the $F_{\mu \nu}\otimes \psi $ cluster from Table~1 could
be explained in two different ways. The first one is, that those
states are difficult to be excited in the direct $\pi N$ channel,
since no totally antisymmetric tensor can be constructed for a
pseudoscalar meson. Only in case the $\pi N N$ vertex would partially
proceed via intermediate $a_1 $ meson states, exciting $F_{\mu
\nu}\otimes \psi$ could become possible. In $\gamma N$ excitations,
however, such states, if existing at all, could become accessible to
measurements. As a second version, one may entertain the possibility,
that the $\{m,0\}\oplus\{0,m\}$ hyperquark configurations in
Eq.~(\ref{h_5}) are of the unusual BWW type. In such a case, the
charge and parity conjugation properties of the BWW hyperquarks will
not any longer match with those of the
$\big\{{k\over2},{k\over2}\big\}$ bosons of usual type, and parity
conservation through strong interaction could not any longer be
guaranteed. Baryons with BWW hyperquarks, would decouple from the
standard nucleon excitation channels and be inaccessible to
measurements running presently.

\section{Mass Formula for the Rarita-Schwinger Clusters}
In this section, we shall argue that the algebra of the degeneracy
group from Eq.~(\ref{Lor_sp}) is also partly the spectrum generating 
algebra. Indeed, the reported mass averages of the resonances from the 
RS multiplets with $l=1,3$, and $5$ are well described by means of the
following simple empirical recursive relation:
\begin{eqnarray}	
M_{\sigma'} - M_{\sigma} = m_1\bigg({1\over {\sigma^2}} - 
{1\over {(\sigma')^2}}\bigg) + {1\over 2} m_2
\bigg({{\sigma '^2-1}\over 2}- {{\sigma ^2 -1}\over 2}\bigg)\, ,
\label{Balmer_ser}
\end{eqnarray}
where, again, $\sigma=k+1$.

The two mass parameters take the values $m_1=600$~MeV, and
$m_2=70$~MeV, respectively. The first term on the r.h.s. in
Eq.~(\ref{Balmer_ser}) is the typical difference between the energies
of two single particle states of principal quantum
numbers $\sigma $, and $\sigma '$, respectively, occupied by a particle 
with mass $m$ moving in a Coulomb-like potential of strength $\alpha_C$ 
with $m_1=\alpha_C^2m/2\hbar^2$. The term
\begin{equation}	
{{\sigma ^2-1}\over 2}=2 {k\over 2}\bigg({k\over 2}+1\bigg)\, ,
\quad \mbox{with} \quad k=\sigma -1 \, ,
\label{I(I+1)_O(4)}
\end{equation}
in Eq.~(\ref{Balmer_ser}) is the generalization of the
three-dimensional $j(j+1)$ rule (with $j=k/2$) to four Euclidean
dimensions \cite{deAlf} and describes a generalized O(4) {\it
rotational band\/}. The parameter $1/m_2=2$, 82~fm corresponds to the
moment of inertia ${\cal J}=2/5 MR^2$ of some ``effective'' rigid-body
resonance with mass $M =1085$~MeV and a radius $R=1$, 13~fm.
Therefore, the energy spectrum in Eq.~(\ref{Balmer_ser}) can be
considered to emerge from a quark--${\cal C}$-hyperquark model with a
Coulomb-like potential (${\cal H}_{\rm Coul}$) and a four-dimensional
rigid rotator ($T_{\rm rot}^{(4)}$). The corresponding Hamiltonian
${\cal H}^{\rm QHM}$ that is diagonal in the basis of the O(4)
harmonics is given by
\begin{eqnarray}	
{\cal H}^{\rm QHM} &=&  {\cal H}_{\rm Coul} + 
T_{\rm rot}^{(4)}  = 
- {\alpha_C\over {|{\bf r}|}} +
{{1}\over {2{\cal J}}} F^2\, .
\label{Hamlit}
\end{eqnarray}

\begin{figure}[htbp]				
\centerline{\psfig{figure=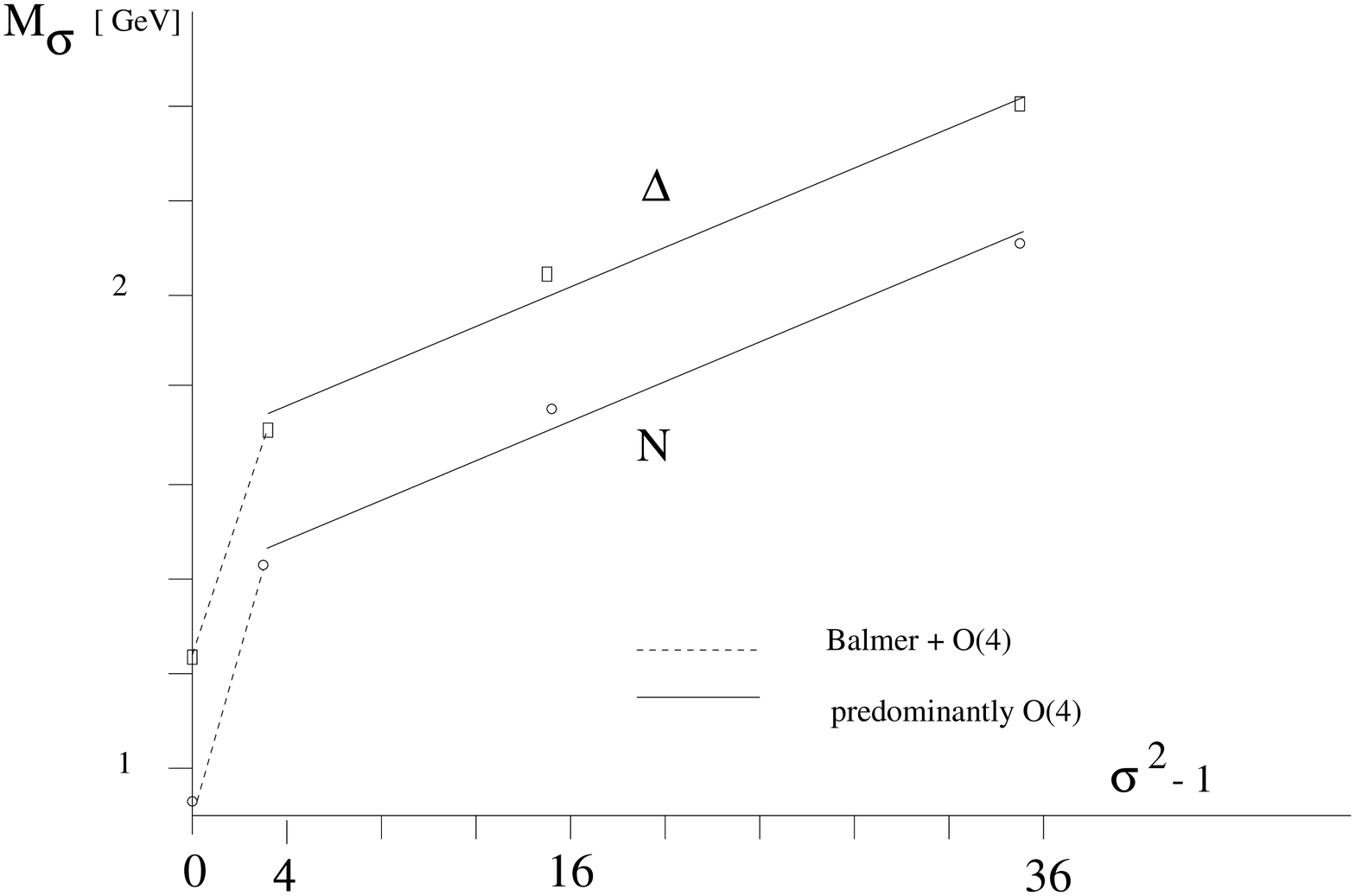,width=10cm}}
\vspace{0.1cm}
{\small Fig.~3\hspace{0.2cm}O(4) rotational bands of nucleon $(N)$ 
and $\Delta$
excitations as modified by the Balmer-series like term in Eq.~(12).}
\end{figure}

\noindent 
Here, ${1\over {2{\cal J}}}F^2$ denotes a rigid rotator in
four Euclidean dimensions as \hbox{associated} with a collective effect
there. Note that while the splitting between the Coulomb-like states
decreases with increasing  $\sigma $, the
difference between the energies of the rotational states increases
linearly with $\sigma $ so that the net effect is an approximate
equidistancy of the baryon cluster levels (see Fig.~3). In extending
the Hamiltonian in Eq.~(\ref{Hamlit}) to include O(4) violating terms
such like $\sim l\cdot s$, $\sim l^2 $, or, $\sim {\bf r}$ (to account
for the confinement), and introducing different moments of inertia
${\cal J}_i$ (to account for possible deformation effects), the O(3)
splitting of the O(4) clusters can be studied along the line of the
collective models of nuclear structure.\cite{EisenbGr}

\section{Summary and Discussion}
We showed that the nucleon and $\Delta $ resonances, instead of being
uniformly distributed in mass, as naively expected on the basis of a
$3q$-Hilbert space without degeneracy, form well pronounced spin- and
parity-clusters. To be specific, we showed that all the reported
nonstrange nucleon resonances with masses below 2500~MeV up to the one
``missing'' state $F_{17}$ expected to occur around 1700~MeV, and the
four more ``missing'' states $H_{1, 11}$, $P_{31}$, $P_{33}$, and
$D_{33}$ states with masses above 2000~MeV, are accommodated by the RS
multispin-parity fields $\psi_\mu$, $\psi_{\mu_1\mu_2\mu_3}$, and
$\psi_{\mu_1\mu_2\cdots \mu_5}$, or, in equivalent notations, the
$2_{2I,+}$, $4_{2I,-}$, and $6_{2I,-}$ Lorentz clusters. This
structure of the observed baryon excitations suggests using a two-body
Hilbert space containing a fundamental point like O$(1,3)$ (or,
Wick-rotated O(4)) bosonic degree of freedom and a fundamental
fermionic one (quark) rather than three independent quark d.o.f. In
this way Lorentz multiplets can appear as bound states. The
``missing'' resonances felt into the $3_{2I,-}$, $3_{2I,+}$, and
$5_{2I,+}$ associated in turn with the RS fields
$A_{\mu_1\mu_2}\otimes \psi$, $V_{\mu_1\mu_2}\otimes \psi$, and
$V_{\mu_1\cdots \mu_4}\otimes \psi$ from Table~1. Also states with
$\{m,0\} \oplus \{0,m\}$ hyperquarks, in our language, are
``missing.'' While the absence of the ``missing'' RS clusters from the
spectra can be interpreted in terms of exclusion of certain type of
parity through the mode of chiral symmetry realization in the
different excitation domains, $[\{m,0\}\oplus\{0,m\}]\otimes\psi$ can
either be excited by photons, or, suppressed, if they turn out to be
of the unusual BWW type. In the idealized case, the 
${\cal C}$-hyperquark--quark system decouples completely from the 
remaining $3q$-configurations and the resonances ``missing'' for the 
completeness of the nonstrange baryons are only five (see Sec.~2).
In reality, some of those couplings may not \hbox{vanish} and few more 
``missing'' resonances can show up.
The masses of the RS-clusters and their spacing were shown to follow 
O(4) rotational bands slightly modified by a Balmer-like term.
The collective character of the baryon excitations illustrated in Fig.~(3)
may indicate that one should not expect to find the complete
$3q$-Hilbert space realized in baryon spectra.

Finally, some remarks on how the model suggested here is related to
QCD --- the gauge theory of strong interaction, are in place. First of
all, the $SU(2)_I\otimes O(1,3)_{ls}$ symmetry considered by us is
simultaneously symmetry of the QCD Lagrangian too, which, in being
manifestly Lorentz invariant, is based upon isovector and isosinglet
quark degrees of freedoms. In addition, the convenience of the
quark--diquark picture is independently supported by an observation
recently under debates in the literature. It concerns the fact, that
the relativistic generalization of the total orbital angular momentum,
$L$, of the three constituent quarks, defined as $\int dx^3
\psi^\dagger ({\bf x}\times (-i D))\psi$, where $\psi = uud$, and $D_i
=\partial_i -i g A_i^a T^a$ does not satisfy the orbital angular
momentum algebra. The introduction of diquark correlations, however,
accounts for a solution of this problem (see Ref.~\cite{Doug} for
details). Moreover, the $1/ r$ potential used here in combination
with a four-dimensional rigid rotator reflects the gluon gauge
dynamics of QCD. The Coulomb-like potential was frequently used in
quark models as it emerges out of the one-gluon exchange between
quarks and is relevant for low energy spectroscopy. However, its
contribution to the $3q$-dynamics is of minor use mainly because of
the collapse of the spectra with increasing principal quantum number.
We here balanced out the decrease of the cluster-level spacing at higher
$\sigma $'s by the energy increase of the O(4) rotational states. 
The significance of the $1/r $
potential below $1,5$~GeV reflects the importance of the one-gluon
exchange in that domain, while the four-dimensional rigid rotator
basically determines the baryon spectrum above $1,5$~GeV and may
point onto an increasing role of the nonperturbative multigluon exchange 
at this scale.

\section*{Acknowledgment}
Work supported by CONACyT Mexico and Deutsche Forschungsgemeinschaft
(SFB~443).

\end{document}

\vspace*{0.21truein}


\begin{thebibliography}{99}
\bibitem{Sternberg} S. Sternberg, {\it Group Theory and Physics} 
(Cambridge University Press, Cambridge, 1994).
\bibitem{Part} C. Caso {\it et al.} (Particle Data Group), 
{\it Eur. Phys. J.} {\bf C3}, 109 (1998).
\bibitem{Ki97-98a} M. Kirchbach, {\it Mod. Phys. Lett.} {\bf A12}, 2373
(1997); {\it ibid.} 3177; {\it ibid.} {\bf A13}, 823 (1998); 
{\it Few Body Syst. Suppl.} {\bf 11}, 47 (1999).
\bibitem{Bhaduri} R. K. Bhaduri, {\it Models of the Nucleon}
(Addison-Wesly, Redwood city, USA, 1988), pp.~24--51.
\bibitem{J00J} S. Weinberg, {\it Phys. Rev.} {\bf 133}, B1318 (1964).
\bibitem{EriceZw} G. Velo and D.  Zwanziger, {\it Phys. Rev.}
                  {\bf 186}, 1337 (1969). 
\bibitem{DVA1} D. V. Ahluwalia, PhD Thesis, Texas A\&M University, 1991; 
{\it Dissertation Abstracts International B} {\bf 52}, 4792B (1992); 
D.~V. Ahluwalia and D.~J. Ernst, {\it Int. J. Mod. Phys.} {\bf E2}, 397 
(1993).
\bibitem{DVA2} D. V. Ahluwalia, M.~B. Johnson and T.~Goldman,
{\it  Phys. Lett.} {\bf B316}, 102 (1993);  D.~V. Ahluwalia, 
{\it Int. J. Mod. Phys.} {\bf A11}, 1885 (1996).
\bibitem{Moshinsky} M. Moshinsky, A. G. Nikitin, A. Sharma and Yu.~F. 
Smirnov, {\it Revista Mexicana de Fisica} {\bf 44} Suppl.~2, 1 (1998).
\bibitem{KiNo} Y. S. Kim and M. E. Noz, {\it Theory and Application of 
the Poincar\'e Group} (D. Reidel Publ. Comp., Dordrecht, 1986).
\bibitem{Kruglov86} S. I. Kruglov and B. I. Strashev, {\it Dokladi Akad. 
Nauk BSSR} {\bf XXX}, 418 (1986) (in Russian).
\bibitem{DVA3} D. V. Ahluwalia, {\it Found. Phys. Lett.} {\bf 10}, 
301 (1997).
\bibitem{Anselmino} M. Anselmino and E. Predazzi, {\it Proc. Int. 
Conf. Diquarks 3}, Torino, Oct. 28--30, 1996 (World Scientific, 1997).
\bibitem{Shibuya} T. Shibuya and C. E. Wulfman, {\it Am. J. Phys.} 
{\bf 33}, 570 (1965).
\bibitem{Hel97} C. Hellstern, R. Alkofer, M. Oettel and H. Reinhardt,
{\it Nucl. Phys.} {\bf A627}, 679 (1997).
\bibitem{Kus} K. Kusaka, G. Piller, A. W. Thomas and A. G. Williams, 
{\it Phys. Rev.} {\bf D55}, 5299 (1997).
\bibitem{deAlf} V. De Alfaro, S. Fubini, G. Furlan and G. Rossetti,
{\it Currents in Hadron Physics} (Elsevier, New York, 1973).
\bibitem{EisenbGr} J. M. Eisenberg and W. Greiner,  {\it Nuclear Models, 
Collective and Single-Particle Phenomena} (North Holland, Amsterdam, 1970).
\bibitem{Doug} D. Singleton, {\it Phys. Lett.} {\bf B427}, 155 (1998).
\end{thebibliography}
\end{document}